\documentclass[fontsize=11pt,twoside=semi,numbers=endperiod]{scrartcl}
\KOMAoptions{headinclude=true,footinclude=false,index=toc}
\usepackage[paperwidth=150mm,paperheight=179.35mm,width=110mm,totalheight=165.35mm,includehead,top=4mm,inner=20mm]{geometry}

\usepackage{amsmath}
\usepackage{amssymb}

\usepackage{lmodern} 

\usepackage[ansinew]{inputenc}
\usepackage[T1]{fontenc}
\usepackage[english,ngerman]{babel}
\usepackage{textcomp} 
\usepackage[pdftex]{color,graphicx} %
\usepackage{bbm}
\usepackage{cite} 
\usepackage{ellipsis} 
\usepackage{microtype} 
\usepackage{scrpage2}
\usepackage[%
pdftex,
colorlinks,citecolor=blue,filecolor=blue,linkcolor=blue,urlcolor=blue,%
bookmarksnumbered=true,
pdfpagemode=UseNone,
pdffitwindow=true,%
pdfstartview={},
pdfpagelayout=SinglePage,
plainpages=false,
pdfpagelabels=true,
]
{hyperref}
\hypersetup{%
    pdfauthor={The Astrophysical Institute, Neunhof},
	pdfsubject={Can the zero-point energy of quantum fields be measured, or is it just an artifact of the theory?},
	pdfkeywords={vacuum,zero-point energy,quantum field theory,cosmological constant, Casimir-effect},
	pdftitle={The Casimir-Effect: No Manifestation of Zero-Point Energy}}

\setkomafont{disposition}{\rmfamily} 
\addtokomafont{section}{\large\bfseries} 
\addtokomafont{subsection}{\normalsize \mdseries \itshape} 
\addtokomafont{pagenumber}{\LARGE}
\addtokomafont{caption}{\small}

\deffootnote[1em]{1em}{1em}{\textsuperscript{\thefootnotemark}\ }

\newcommand{\bz}[1]{\nolinebreak\hspace{0em}\nolinebreak{}#1\hspace{0em}}

\newcommand{\ggcite}[1]{\nolinebreak\mbox{\hspace{0.2em}\cite{#1}}\hspace{.05em}} 
\newcommand{\ggoptcite}[2][]{\nolinebreak\mbox{\hspace{0.2em}\cite[#1]{#2}}\hspace{.05em}}

\clearscrheadfoot 
\rohead[\vspace{-2mm}\par\pagemark ]{\vspace{-2mm}\par\pagemark}
\lehead[\vspace{-2mm}\par\pagemark ]{\vspace{-2mm}\par\pagemark}
\rehead{ }
\lohead{ }

\begin{document}
\selectlanguage{english}
\thispagestyle{empty}
\raggedbottom  
\vspace*{3mm}
\centerline{\bfseries\Large The Casimir-Effect:}
\vspace*{.5mm}
\centerline{\bfseries\Large No Manifestation of Zero-Point Energy}
\vspace*{2mm}
\centerline{{\bfseries Gerold Gr\"{u}ndler}\,\footnote{e-mail:\ gerold.gruendler@astrophys-neunhof.de}}
\vspace*{1mm}
\centerline{\small Astrophysical Institute Neunhof, N\"{u}rnberg, Germany} 
\vspace*{3mm}
\noindent\parbox{\textwidth}{\small The attractive force between metallic surfaces, predicted by Casimir in 1948, seems to indicate the physical existence and measurability of the quantized electromagnetic field's zero\bz{-}point energy. It is shown in this article, that the measurements of that force do not confirm Casimir's model, but in fact disprove it's foundational assumption that metal plates may be represented in the theory by quantum\bz{-}field\bz{-}theoretical boundaries. The consequences for the cosmological constant problem are discussed.} 
\vspace*{1mm}

\section{Is the zero-point energy of quantized fields observable?}
General relativity theory (GRT), and the relativistic quantum field theories (QFT) of the standard model of elementary particles, are describing all experimental observations with impressive accuracy --- as long as GRT or QFT are used separately. But as soon as one tries to combine these successful theories, serious problems turn up. One of the most spectacular examples of such incompatibility has been dubbed ``the cosmological constant problem''. A by now classical review article on that issue has been compiled by Weinberg\ggcite{Weinb:KosmConst}. See the article by Li \mbox{\em et.\,al.}\ggcite{Miao:darkenergy} for an updated review. In short, the cosmological constant problem manifests itself as follows: 

According to the field equation\pagebreak  
\begin{align}
R_{\mu\nu}-\frac{R}{2}\, g_{\mu\nu}+\Lambda\, g_{\mu\nu}=-\frac{8\pi G}{c^4}\,T_{\mu\nu} \label{lkdynjbgf}
\end{align}
of GRT, the curvature of space\bz{-}time, represented by the Ricci\bz{-}tensor $(R_{\mu\nu})$ and it's contraction $R$, is proportional to the energydensity\bz{-}stress\bz{-}tensor $(T_{\mu\nu})$, which again is determined by the energy density and the momentum density of all fields contained within space\bz{-}time, \mbox{i.\,e.} of all fields with exception of the metric field $(g_{\mu\nu})$. 

Upon canonical quantization of any classical continuous field, the energydensity\bz{-}stress\bz{-}tensor of that field will diverge. For example, canonical quantization of the classical electromagnetic field results into the Hamilton operator 
\begin{subequations}\begin{align}
H&=\int\!\text{d}^3x\;T_{00}\notag\\ 
&=\sum _{\boldsymbol{k}}\;\sum _{v=1}^2\frac{\hbar c\,|\boldsymbol{k}|}{2}
\Big( a^{{(v)}+}_{\boldsymbol{k}}a^{(v)}_{\boldsymbol{k}}+a^{(v)}_{\boldsymbol{k}}a^{{(v)}+}_{\boldsymbol{k}}\Big) \label{owghrasga} \\
&=\sum _{\boldsymbol{k}}\;\sum _{v=1}^2\hbar c\,|\boldsymbol{k}|\Big( a^{{(v)}+}_{\boldsymbol{k}}a^{(v)}_{\boldsymbol{k}}
+\frac{1}{2}\underbrace{[a^{(v)}_{\boldsymbol{k}},a^{{(v)}+}_{\boldsymbol{k}}]}_{1}\Big)\ ,\label{owghrasgb} 
\end{align}
with $a^{{(v)}+}_{\boldsymbol{k}}$ and $a^{(v)}_{\boldsymbol{k}}$ being the creation\bz{-} and annihilation\bz{-}operators respectively of photons with wavenumber $\boldsymbol{k}$ and polarization $v$. The integration is over the complete normalization volume, and the summation is running over all of the infinitely many wavenumbers $\boldsymbol{k}$, which are compatible with the normalization volume (if an infinite normalization volume is chosen, the sum over $\boldsymbol{k}$ is replaced by an integral over $\boldsymbol{k}$). Due to the commutator, the energy is infinite. The waves described by the second term in \eqref{owghrasgb} are the zero\bz{-}point\bz{-}oscillations, and their energy is the zero\bz{-}point\bz{-}energy, of the quantized electromagnetic field.

The observed curvature of intergalactic space is close to zero \cite{Bennet:WMAP9y}, suggesting that either 
\newline{}(a)\;{}the cosmological constant $\Lambda $ in \eqref{lkdynjbgf} should be adjusted, to compensate the zero\bz{-}point\bz{-}energy of the quantum fields, or that
\newline{}(b)\;{}the zero\bz{-}point\bz{-}energy of the quantum fields should be considered as a strange artifact of the theory without analog in observable reality, and therefore be removed somehow from QFT. 

To avoid misunderstandings, we note that the zero\bz{-}point oscillations of quantum fields with only a finite number of degrees of freedom, \mbox{e.\,g.} the zero\bz{-}point oscillations of the phonon fields of molecules and solids, have been experimentally confirmed since almost a century \cite{Mulliken:bospectr,James:NaClzeropten}. But what we are exclusively discussing in this article is the zero\bz{-}point energy of elementary quantum fields with infinitely many degrees of freedom. 

Alternative (a) calls for a fine\bz{-}tuning of the cosmological constant $\Lambda $ with an accuracy of many dozens of decimal digits, the exact number depending on the method applied for regularization of the diverging term in \eqref{owghrasgb}. Therefore this solution --- though being completely correct under purely formal criteria --- does seem to be quite ``unnatural'', and is not considered acceptable by many scientists. 

On first sight, there seem to be less objections against alternative (b). In pure quantum\bz{-}field\bz{-}theoretical computations (neglecting gravity), only energy differences matter, but not absolute energy values. Therefore the offset of an infinitely large zero\bz{-}point energy is merely a tiresome ballast without discernible functionality. To get rid of that offset, normal order is often applied as an ad\bz{-}hoc measure. It means, that in \eqref{owghrasga} all creation operators are shifted left, and all annihilation operators are shifted right, under disregard of their commutation relations! Thus the Hamilton\bz{-}operator becomes 
\begin{align}
H=\sum _{\boldsymbol{k}}\;\sum _{v=1}^2\,\hbar c\,|\boldsymbol{k}|\, a^{{(v)}+}_{\boldsymbol{k}}a^{(v)}_{\boldsymbol{k}}\ ,\label{owghrasgc}  
\end{align}\end{subequations}
and the infinite energy offset has disappeared. 

As the results of quantum\bz{-}field\bz{-}theoretical computations are not changed, if normal order is applied to the Hamilton operator, and as no gravitational effect of the zero\bz{-}point\bz{-}energy of quantum fields is observed, one might very well ask whether that zero\bz{-}point energy does exist at all. Words like ``existence'' or ``reality'' in this context of course mean the question, whether the zero\bz{-}point energy is observable and can be tested experimentally. 

\section{The Casimir-effect} 
A possible method to observe the quantized electromagnetic field's zero\bz{-}point energy --- and actually the only method proposed until today, besides the missing gravitational effect --- has been suggested by Casimir\ggcite{Casimir:caseffect} in 1948. 
\begin{figure}[!b]\begin{center}
\includegraphics[scale=0.8]{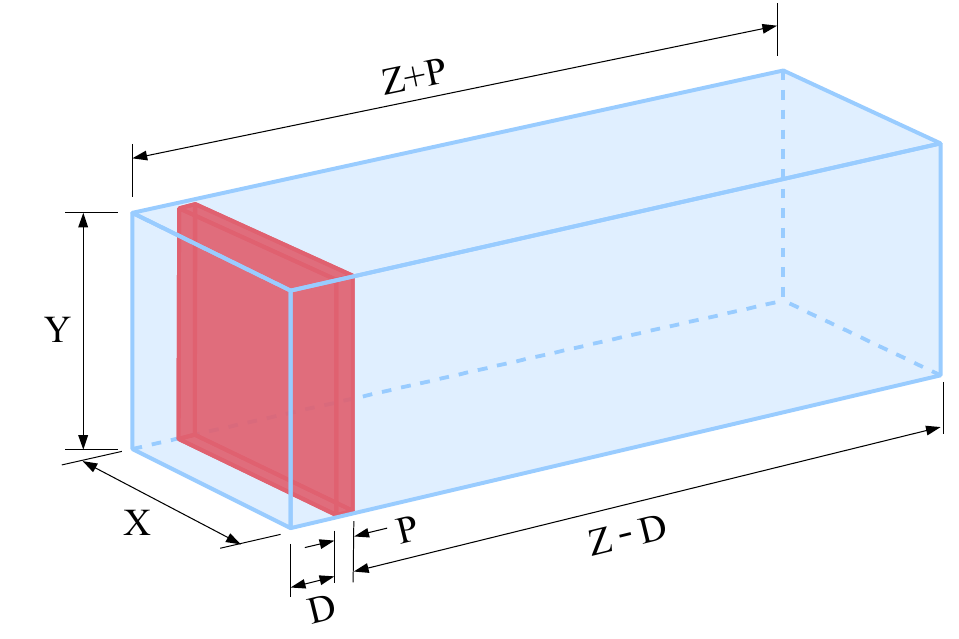} 
\caption{Cavity resonator with movable plate}\label{fig:hohll}
\end{center}\end{figure}
Casimir considered a resonator as sketched in figure \ref{fig:hohll}. The rectangular cavity's size is $X\times Y\times (Z+P)$. Inside the cavity there is a plate of thickness $P$, which is aligned parallel to the cavity's $XY$\bz{-}face and movable in $Z$\bz{-}direction. The plate's distance from one side wall of the cavity is $D$, it's distance from the opposite side wall is $Z-D$. 

The resonance spectra of the left and right cavities are discrete. The wavenumbers are 
\begin{align}
&k_{rst}=\sqrt{\Big(\frac{r\pi}{X}\Big) ^2+\Big(\frac{s\pi}{Y}\Big) ^2+\Big(\frac{t\pi}{A}\Big) ^2}\notag\\ 
&\text{with }r,s,t\in\mathbbm{N}\ ,\label{ifjgfjdga}\\ 
&\mbox{\parbox{.9\linewidth}{There are 2 modes each with $r,s,t=1,2,3,\dots $ and 1 mode each with one of the indices 0 and the both other indices $1,2,3,\dots $ \cite[chap.\,D.II.2.b.]{Gundlach:hochfrtech}}}\notag 
\end{align}
with $A=D$ for the left cavity and $A=Z-D$ for the right cavity. Casimir identified the plate and the walls of the cavity with the boundaries of the normalization volume of quantum\bz{-}electrodynamics\,=\,QED. Therefore he considered this equation not only valid for photons, but as well for the zero\bz{-}point oscillations of the quantized electromagnetic field. According to this point of view, long\bz{-}wavelength zero\bz{-}point oscillations, which don't fit into the cavities, can not evolve in the respective volumes. Casimir computed the zero\bz{-}point energy $U_{\text{left}}$ enclosed in the left cavity, and the zero\bz{-}point energy $U_{\text{right}}$ enclosed in the right cavity. A detailed account of Casimir's computation can be found elsewhere\mbox{\ggoptcite[sect.\,4]{apin:se08011}.} Both $U_{\text{left}}$ and $U_{\text{right}}$ are functions of $D$, and both are of course infinite, as infinitely many short\bz{-}wavelength zero\bz{-}point oscillation modes fit into the cavities. But the derivative 
\begin{align}
F_{\text{Casimir}}&\equiv\frac{\text{d}(U_{\text{left}}+U_{\text{right}})}{\text{d}D} \label{osghsdhgdd}
\end{align}
is finite! Under the assumption, that the cavity walls and the movable plate perfectly reflect electromagnetic radiation of arbitrary frequencies, and assuming $Z\gg D$, Casimir found a surprisingly simple result: 
\begin{align}
F_{\text{Casimir}}=-\,\frac{\pi ^2\hbar c}{240}\,\frac{XY}{D^4}
=-1.3\cdot 10^{-9}\text{N}\cdot\frac{XY/\text{mm}^2}{D^4/\mu\text{m}^4}  \label{caskraft}
\end{align}
This is a small, but measurable force, which is pushing the movable plate towards the nearer cavity wall. It has by now been measured many times, and the approximate correctness of equation \eqref{caskraft} has been confirmed \cite{Lambrecht:caseff2}. It is no surprise, that the experimental confirmation is only approximate but not exact, because Casimir derived \eqref{caskraft} not for real metal plates, but for perfectly reflecting plates, \mbox{i.\,e.} boundaries in the terminology of quantum field theory. The important differences in the physical concepts of real metals and boundaries will be discussed in the next section. Based on the approximate experimental confirmations of \eqref{caskraft}, the attractive forces between metallic plates have been declared to be ``physical manifestations of zero\bz{-}point energy''\ggcite{Milton:caseffbuch}. 

These observations, however, do not conclusively prove the reality of zero\bz{-}point energy, because an alternative explanation for the same observations is available, which does not refer at all to zero\bz{-}point energy: Lifshitz\ggcite{Lifshitz:vanderwaal} and Dzyaloshinskii, Lifshitz, and Pitaevskii\ggcite{Dzyaloshinskii:vanderwaal} have computed the retarded van der Waals\bz{-}force, which is acting between two infinitely extended half\bz{-}spaces with relative dielectric constants $\epsilon _1$ and $\epsilon _3$, while the gap between them is filled with a material with relative dielectric constant $\epsilon _2$. Schwinger, DeRaad, and Milton\ggcite{Schwinger:casindie} reproduced and confirmed the results of Lifshitz \mbox{\em et.\,al.\,.} They also considered the limit $\epsilon _1=\epsilon _3\rightarrow\infty$, $\epsilon _2\rightarrow 1$, \mbox{i.\,e.}\ the limit of two metal plates with infinite conductivity, separated by a vacuum gap. It turned out that the formula of Lifshitz \mbox{\em et.\,al.} simplifies in this limit to the Casimir\bz{-}force \eqref{caskraft}. 

The question ``does the observed Casimir\bz{-}force prove the observable existence of zero\bz{-}point energy?'' at first sight seems not to be answered by the theory with a clear\bz{-}cut {\small YES} or {\small NO}, because there are two different theoretical concepts, one of them indicating {\small YES}, and the other indicating {\small NO}. But closer scrutiny reveals, that the answer definitely is {\small NO}. Jaffe\,\cite{Jaffe:Casimir} remarked, that Casimir's assumption of perfect reflectivity of the metal plates is equivalent to taking the limit $\alpha\rightarrow\infty $, with $\alpha $ being the QED coupling constant, and thus is obscuring the true nature of the interaction between the electromagnetic field and the charged matter\bz{-}fields constituting the metallic plates. While this criticism certainly is justified and pointing into the right direction, it is missing --- or at least not explicitly naming --- the essential point: Only the Lifshitz model is compatible with the results of measurements, while Casimir's model is refuted by experimental evidence. 

The two explanations of the Casimir\bz{-}force are predicting similar, but not identical values of that force: The model, which is based on van\,der\,Waals\bz{-}forces, can match the measurement results exactly if parameters like the complex dielectric constant $\epsilon (\omega )=\epsilon '(\omega )+i\epsilon ''(\omega )$ as a function of photon frequency $\omega $ are adjusted \cite{Lambrecht:caseff2}. In contrast, in Casimir's model there are no adjustable parameters, see his equation \eqref{caskraft}. And it is an essential feature of Casimir's model, that no adjustable parameters like \mbox{e.\,g.} a reflection coefficient $<1$ can be introduced without \emph{complete demolition} of the model. This assertion will be proved in the next section. 

Only the model of Lifshitz \mbox{\em et.\,al.} can stand the confrontation with the results of measurements, while the results derived from Casimir's model differ typically by about 10 to \mbox{20\,\%} at a plate distance of \mbox{$1\,\mu $m} from experimental observations \cite[sect.\,5.2]{Bordag:Casimir}. Casimir's explanation of the Casimir\bz{-}force is disproved by the experiments, because a significant discrepancy of about 10 to \mbox{20\,\%} between theory and experiments, which \emph{can not be eliminated due to improvement of the model}, is about 10 to \mbox{20\,\%} to much. 

\section{Metals and boundaries} 
At the outset of any quantum field theory, a normalization volume (which may be finite or infinite) must be fixed, and well\bz{-}defined boundary conditions must be imposed onto the field at the boundaries. A simple choice is for example a Dirichlet\bz{-}type boundary condition, requiring that the field amplitude must be zero at the boundary. Another, often more convenient choice is a periodic (Cauchy\bz{-}type) boundary condition, requiring that the value and the derivative of the field at one point of the boundary must at any time be identical to the value and the derivative of the field at the opposite point of the boundary. Either of theses conditions makes sure, that the norm 
\begin{align}
\langle s|s\rangle =N\in\mathbbm{R}\quad ,\quad 0<|N|<\infty\label{sdjhgkfda}
\end{align}
of any state\bz{-}function $|s\rangle $ has a well\bz{-}defined value $N$ which can be normalized to unity, and --- most important! --- which is constant. This means, that either no probability density assigned to the state $|s\rangle $ can penetrate through the boundaries (Dirichlet\bz{-}type boundary condition), or that probability density flowing out of the normalization volume at one spot of the boundary is exactly compensated by probability density flowing into the normalization volume at the opposite spot of the boundary (Cauchy\bz{-}type boundary condition). 

The resonance spectrum \eqref{ifjgfjdga} is enforced by the boundary condition 
\begin{align}
\boldsymbol{E}_{\text{tangential}}(\text{surface})=\boldsymbol{H}_{\text{normal}}(\text{surface})=0\label{lsgjgddg}
\end{align}
onto the electrical amplitude $\boldsymbol{E}$ and the magnetizing amplitude $\boldsymbol{H}$ of the electromagnetic field at any spot of the surface of the cavity walls and the surface of the plate. No real metal can enforce this condition onto the field, but only an ideal material which is reflecting \mbox{100\,\%} of impinging radiation at any frequency. 

If a photon impinges onto a plate made of real metal, then it may be reflected, or it may be absorbed, or it may be transmitted. For example, good electrical conductors like copper or gold reflect most long\bz{-}wavelength sub\bz{-}infrared photons, reflect about half and absorb about half of optical photons, and are almost transparent for short\bz{-}wavelength X\bz{-}ray photons. Note that at least a small part of the impinging photons are absorbed by any metal at almost any frequency. Only superconductors absorb strictly no photons of sufficiently long wavelength, but even they absorb photons of infrared and shorter wavelengths. If one wants to compute the resonance spectrum of a cavity made from real metal, one therefore needs to relax condition \eqref{lsgjgddg}, and allow for absorption by and transmission through the cavity walls in particular with regard to high\bz{-}frequency radiation. This results into damping and broadening of resonance modes. 

In contrast, the well\bz{-}defined boundary conditions of the normalization volume must not be relaxed under any circumstances, because a damped norm like 
\begin{align}
\langle s|s\rangle =Ne^{-\gamma t}\quad ,\quad &N,\gamma ,t\in\mathbbm{R}\ ,\notag\\ 
0<&|N|,\gamma ,t<\infty\ , \label{lsdfhgjmhg}
\end{align}
with $\gamma $ being some damping parameter, and $t$ being time, would not be a reasonable extension of \eqref{sdjhgkfda}, but a contradiction to the basic tenets of quantum field theory. If for example $|s\rangle $ is a state, in which exactly one photon is excited, then of course this photon may disappear after some time due to interaction with matter. But \eqref{lsdfhgjmhg} would imply that the photon would little by little disappear even without any interaction, \mbox{i.\,e.} somehow slip out from the normalization volume. 

To exclude the senseless result \eqref{lsdfhgjmhg}, boundaries must not be merely approximate boundaries, which allow for damping due to dissipation of probability density. Even a good approximation would not be sufficient. Only perfect boundaries are good enough, because only perfect boundaries can guarantee that probability density is conserved and that the norm of state\bz{-}functions is constant. The boundaries of any quantum field theory are mathematical, not tangible entities, and no real metal can replace the boundaries of QED. 

Fields and boundaries exhaust the inventory of quantum field theory. Every entity of physical reality must be represented in QFT either by a boundary or by a field. If realistic parameters are assigned to the cavity walls and the plate, then they can not be described as boundaries but must be described as material fields, like conduction band electrons, or crystal ions, or Cooper pairs in case of superconductors, or whatever types of charged matter fields, which can couple to the electromagnetic field. As the electromagnetic field according to QED is nothing but the gauge field of charged matter fields, the interaction between the matter fields and their gauge field is uniquely defined: Photons, described by state functions like $|\boldsymbol{k},v\rangle =a^{{(v)}+}_{\boldsymbol{k}}|0\rangle $ or linear combinations of such state functions, are the electromagnetic field's quanta, which can couple to charged matter fields. If the electromagnetic field is in the the vacuum\bz{-}state $|0\rangle $, in which no photons are excited but only the electromagnetic field's zero\bz{-}point oscillations exist, then it does not couple to any field, but only to boundaries. As soon as the quality of boundaries is no more ascribed to the walls of the cavity and to the plate, the electromagnetic field's zero\bz{-}point oscillations vanish from the picture. If the metals are represented in the theory not by boundaries but by matter fields, then the interaction between the metals and the electromagnetic field is not affected, if zero\bz{-}point energy is skipped from the theory due to normal order of the Hamilton operator. 

The geometry of the boundaries can be chosen in quantum field theory to a large extend at will. But of course the boundaries must enclose all parts of the physical system to be described. If the resonator depicted in \mbox{fig.\:\ref{fig:hohll}} shall be described with walls and plate made from real metals, then the boundaries must enclose, besides the cavity space, both the walls and the plate, because photons may be absorbed by them or may penetrate through them. The volume enclosed by the boundaries must be larger than the cavity volume, and consequently the spectrum of photon\bz{-}wavenumbers in this setup of QED will not be identical to the spectrum of the resonator's resonance wavenumbers. In particular, the spectrum of the zero\bz{-}point oscillations, which is determined by the geometry of the QED boundaries, is in this setup not related to the spectrum of the resonator's resonance wavenumbers, because the resonator walls must be described as matter fields, which do not interact with the electromagnetic field's zero\bz{-}point oscillations. 

Side note: From the field\bz{-}theoretical point of view it is obvious, that only boundaries, but not real metal plates with finite conductivity, can shape the spectrum of zero\bz{-}point oscillations. In the mid of the second page of his article\ggcite{Casimir:caseffect}, Casimir made a quite strange remark to the contrary. There he pointed out that most zero\bz{-}point oscillations of very short wavelength (\mbox{e.\,g.} of X-ray wavelength) would penetrate through metals, while most long\bz{-}wavelength zero\bz{-}point oscillations would be reflected. \mbox{I.\,e.} he assumed that the reflection spectra of metals are similar (if not identical) for photons and for zero\bz{-}point oscillations. The present author undertook the tedious task, to evaluate the consequences of that assumption\mbox{\ggcite{apin:se93327}.} Not surprisingly it turned out that Casimir's strange assumption is leading to results which are contradicting the experimental observations. Note that Casimir's final result \eqref{caskraft} is not affected by that strange assumption, because he achieved that result for a model, in which the plates are represented by boundaries but not by metals. Therefore \eqref{caskraft} is independent of any considerations on the interaction of zero\bz{-}point oscillations with real metals. 

Idealized conditions can be considered approximations to realistic conditions, if the conditions can (at least theoretically) be gradually changed from the idealized to the realistic case. For example, one could approach a realistic scenario by first assuming no interaction between matter and radiation, \mbox{i.\,e.} setting the QED coupling constant $\alpha $ to zero, and then step by step improve the approximation by increasing $\alpha $ gradually to $\alpha\approx 1/137$. But no gradual transition is possible from a boundary to a matter field. A boundary either is a perfect boundary, or it is no boundary at all. If one tries to gradually improve the approximation of Casimir's model, in which metal plates are represented by boundaries, \mbox{i.\,e.} in which $\alpha =\infty $ is assumed, then the quality of boundaries is abruptly removed from the plates and thus \emph{the foundations of the model are completely destroyed} as soon as one decreases $\alpha $ to a finite (even if arbitrary high) value. 

Thus the significant differences between the experimental observations and the predictions of Casimir's model, which have been mentioned at the end of the previous section, can not be eliminated nor diminished by assigning to the boundaries the reflectivity of metals. Casimir's model of the Casimir\bz{-}force, in which metal plates are represented by boundaries, is an ingenious theoretical construction, but by experimental evidence ruled out as a correct description of reality. Therefore the observed Casimir\bz{-}force does not indicate the physical existence of zero\bz{-}point energy. 

\section{Conclusions}
The findings presented in this article may be helpful to avoid blind alleys, and to stir the search for a solution of the cosmological constant problem into the right direction. The essential facts are: Firstly, no gravitational effect, caused by the zero\bz{-}point energy of quantum fields, has been observed. Secondly, the results of QFT --- including all of the impressive achievements of QED like Lamb\bz{-}shift, electron g\bz{-}factor, hydrogen hyperfine\bz{-}splitting, and so on --- are not compromised, if zero\bz{-}point energy is skipped from the theory due to application of normal order \eqref{owghrasgc} to the Hamilton operator. Thirdly, as shown in this article, the observed Casimir\bz{-}force does definitely not prove the reality of zero\bz{-}point energy. 

In total, no experimental evidence at all is indicating the measurable, observable existence of the zero\bz{-}point energy of elementary quantum fields. Therefore, instead of renormalizing the cosmological constant $\Lambda $ (or even modifying GRT), it is certainly more promising to approach the problem directly at it's root, \mbox{i.\,e.} to somehow remove zero\bz{-}point energy from QFT. The crude measure of normal order is not an acceptable solution, as the disregard of the non\bz{-}commutative operator algebra is irreconcilable with the basic principles of QFT. Unfortunately, at this moment I cannot offer a better solution. 

\section*{Acknowledgments}
This work gained much from innumerable helpful discussions on the mysterious quantum vacuum with my colleagues \mbox{V.\,I.\,Nachtmann} and \mbox{O.\,S.\,ter\,Haas}. 

\renewcommand{\bibname}{References}
\flushleft{\bibliography{../../gg}} 
\end{document}